# Dilepton Production at SPS Energies*

W. Cassing, W. Ehehalt and C. M. Ko†

Institut für Theoretische Physik, Universität Giessen

D-35392 Giessen, Germany

August 4, 1995

## Abstract

We present a nonperturbative dynamical study of $e^+e^-$ production in proton-nucleus and nucleus-nucleus collisions at SPS energies on the basis of a covariant transport approach. For p + A reactions the dilepton yield for invariant masses m $\leq$ 1.2 GeV is found to be dominated by the decays of the $\eta, \rho, \omega$ and $\Phi$ mesons in line with the findings of the CERES collaboration. For S + Au collisions at 200 GeV/A the dilepton yield is, however, dominated by $\pi^+\pi^-$ annihilation due to the high pion densities achieved. Whereas for 'free' meson masses and form factors the experimental cross section is slightly underestimated for 0.3 GeV $\leq$ m $\leq$ 0.45 GeV, different medium modifications of the $\rho$-meson appear compatible with current CERES data.

---

*Work supported by BMFT and GSI Darmstadt.

†Permanent address: Cyclotron Institute and Physics Department, Texas A&M University, College Station, Texas 77843

During the last few years a major effort in heavy-ion collisions up to bombarding energies of 200 GeV/A has been to probe the properties of hot and dense nuclear matter as well as its phase transition to the quark-gluon plasma. Apart from signatures associated with the quark-gluon degrees of freedom the partial restoration of chiral symmetry at high baryon density is of fundamental interest in its own right. Particle production plays a special role in this context [1] since mesonic and electromagnetic probes, that are not available in the initial stage of the reaction, carry information on the possible transition phase.

Contrary to mesons, electromagnetic signals are particularly well suited for an investigation of the violent phases of a high-energy heavy-ion collision because they can leave the reaction volume essentially undistorted by final-state interactions. Whereas the direct photon signal is overwhelmed by the strong background of meson decay photons, dileptons ($e^+e^-$ and $\mu^+\mu^-$ pairs) are free from such problems. Indeed, dileptons from heavy-ion collisions have been observed by the DLS collaboration at the BEVALAC [2, 3, 4] and by the CERES [5] and HELIOS collaboration [6] at SPS energies.

Furthermore, dileptons can also be used as probes for the coupling of time-like photons to charged hadrons in the nuclear medium. The vector-meson dominance model assumes that this coupling proceeds through virtual $q\bar{q}$ excitations of the QCD vacuum with the quantum numbers of the photon, i.e. through the vector mesons (mainly the $\rho$-meson). Dileptons thus are expected to be sensitive to the properties of the $\rho$-mesons in the medium since the latter predominantly decay still in the dense nuclear medium due to their short lifetime.

In fact, both the CERES and HELIOS collaborations have found an enhancement of dileptons in A + A collisions compared to p + A collisions for invariant masses 0.3 GeV $\leq$ m $\leq$ 0.6 GeV. Studies by Srivastava *et al.* [7] based on a quark-gluon scenario can not account for the excess dileptons. On the other hand, this enhancement has been interpreted by Li *et al.* [8] as a signature for chiral symmetry restoration, more precisely, as a signature for the dropping mass of the $\rho$-meson in the dense medium. However, their analysis was based on an expanding fireball scenario in chemical equilibrium, which validity in S + Au collisions might be questionable. In this paper, we will thus carry out a partially related but extended study based on a nonequilibrium covariant transport model.

In continuation of our work in refs. [9, 10, 11, 12] we study the dynamics of proton-nucleus or nucleus-nucleus reactions by using a coupled set of covariant transport equations with scalar and vector self-energies of all hadrons involved. Explicitly propagated are nucleons, $\Delta$'s, N*(1440), N*(1535) resonances as well as $\pi$'s, $\eta$'s, $\rho$'s, $\omega$'s, $\Phi$'s, kaons and K*'s with their isospin degrees of freedom. For more detailed information on the self-energies employed we refer the reader to ref. [13], where the transport approach HSD[1] is formulated and applied to nucleus-nucleus collisions from SIS to SPS energies.

In this investigation we calculate dilepton production taking into account the contributions from nucleon-nucleon, pion-nucleon and pion-pion bremsstrahlung, the Dalitz-decay of the $\Delta$, N*(1440), $\eta \to \gamma e^+ e^-$ and $\omega \to \pi^0 e^+ e^-$, the direct dilepton decays of the vector mesons $\rho, \omega, \Phi$ as well as $\pi^+ \pi^-$ annihilation. The nucleon-nucleon, pion-nucleon and $\pi\pi$ bremsstrahlung, the Dalitz-decay of the $\Delta, N^*$, $\pi^0$ and $\eta$ are evaluated in the same way as described in refs. [9, 11].

The pion annihilation - which will turn out to play a specific role - proceeds through the $\rho$-meson which decays into a virtual massive photon by vector meson dominance. The cross section is parametrized as in [9, 11, 14] as

$$\sigma^{\pi^+\pi^- \to e^+e^-}(m) = \frac{4\pi}{3}\left(\frac{\alpha}{m}\right)^2 \sqrt{1 - \frac{4m_\pi^2}{m^2}} |F_\pi(m)|^2 , \qquad (1)$$

where the free electromagnetic form factor of the pion is given by

$$|F_\pi(m)|^2 = \frac{m_\rho^4}{(m^2 - m_\rho'^2)^2 + m_\rho^2 \Gamma_\rho^2} . \qquad (2)$$

---
[1]Hadron-String-Dynamics

In eq. (2) $m$ is the dilepton invariant mass, $\alpha$ is the fine structure constant, and

$$m_\rho = 775 MeV, \qquad m'_\rho = 761 MeV, \qquad \Gamma_\rho = 118 MeV.$$

Note that eq.(1) describes the free pion annihilation cross section; possible medium modifications will be discussed below.

For the cross section of the proton-neutron bremsstrahlung we use the phase-space corrected soft-photon approximation which has been shown to be a good approximation to a more fundamental one-boson-exchange calculation [9, 15]. This approximation might be questionable for pion-nucleon and pion-pion bremsstrahlung, but due to their rather small contributions (see below) relative factors of 2-3 will not change the conclusions of our present study. Similar transport models for $e^+e^-$ production are reported in refs. [16, 17, 18, 19, 20].

In addition to our previous calculations we now also include the $\omega$ Dalitz-decay given by [21]:

$$\frac{d\Gamma_{\omega \to \pi^0 e^+ e^-}}{dm} = \frac{\alpha}{3\pi} \frac{\Gamma_{\omega \to \pi^0 \gamma}}{m} \left( (1 + \frac{m^2}{m_\omega^2 - m_\pi^2})^2 - \frac{4 m_\omega^2 m^2}{(m_\omega^2 - m_\pi^2)^2} \right)^{3/2} \times |F_{\omega \to \pi^0 e^+ e^-}(m)|^2, \qquad (3)$$

where the form factor is parametrized as ([21])

$$F_{\omega \to \pi^0 e^+ e^-}(m) = \frac{1}{1 - m^2/\Lambda_s^2} \qquad (4)$$

with

$$\Lambda_s = 0.65 \text{ GeV}. \qquad (5)$$

The singularity for m = $\Lambda_s$ is avoided by introducing numerically a finite width in the denominator which is compatible with the data on $F_\omega$ in ref. [21]. The direct decays of the vector mesons to $e^+e^-$ relative to the total width are taken as $\Gamma_{\Phi \to e^+ e^-}/\Gamma_\Phi^{tot} = 3.09 \times 10^{-4}$, $\Gamma_{\omega \to e^+ e^-}/\Gamma_\omega^{tot} = 7.1 \times 10^{-5}$ and $\Gamma_{\rho \to e^+ e^-}/\Gamma_\rho^{tot} = 4.4 \times 10^{-5}$.

We have calculated the dilepton yields for p + Be and p + Au at 450 GeV, and for S + Au at 200 GeV/A bombarding energy. A comparison with the experimental data of the CERES collaboration [5] is shown in Fig. 1 for p + Be and p + Au including the dilepton acceptance cuts in pseudo-rapidity, i.e. $2.1 \leq \eta \leq 2.65$, a cut of the transverse dilepton momenta for $p_T \geq 0.05$ GeV/c as well as a cut on the opening angle of the dileptons $\Theta \geq 35$ mrad. Also, the experimental mass resolution has been included in evaluating the theoretical mass spectrum. The full solid curves in Fig. 1 display the sum of all individual contributions which is dominated by the decays of the mesons. The bremsstrahlung contributions ($\pi$N, pN) as well as $\pi^+\pi^-$ annihilation are of minor importance for both systems in line with the experimental cocktail analysis in ref. [5].

Before going over to the system S + Au at 200 GeV/A we need to check if the global reaction dynamics are reproduced by our transport calculation. This is obviously the case, as may be extracted from Fig. 2, where we compare our calculated rapidity distribution for protons (dashed line) and $\pi^-$ (solid line) with the experimental data [22, 23]. The proton rapidity spectrum shows a narrow peak at target rapidity ($\approx$ -3.03) which is easily attributed to the spectators from the Au target. The bump at y $\approx$ -2, furthermore, is mainly due to rescattering of target nucleons. Please note that there is no longer any yield at projectile rapidity (y $\approx$ 3.03) which implies that all nucleons from the projectile have undergone inelastic scatterings. The CERES acceptance roughly covers the rapidity regime $-1 \leq y \leq -0.4$ where the pion density is large compared to the proton density.

We now turn to the calculations for $e^+e^-$ production, where the experimental cuts have been employed again. Contrary to the p + Be and p + Au reactions, a cut on the transverse dilepton momenta $p_T \geq 0.2$ GeV/c has been taken in line with the experimental acceptance [5]. The results of our calculation, where no medium effects are incorporated for all mesons, are displayed in Fig. 3 (upper thick solid line) in comparison to the data [5]. The open triangles present the experimental extrapolation for mesonic decays [5] whereas the thick solid line denoted by 'cocktail' is the result of $\eta, \omega, \Phi$ and primary $\rho$ decays from our calculation. Except for a single point at invariant mass m = 0.6 GeV our cocktail analysis coincides with that of the CERES collaboration such that we can attribute the additional yield seen experimentally (full dots) to the additional channels accounted for in our model. In fact, the $\pi^+\pi^-$ annihilation component (dashed line) is found to dominate the dilepton spectrum for m $\geq$ 0.3 GeV very much reminiscent of the situation at BEVALAC or SIS energies [9, 11, 17]. The additional bremsstrahlung contributions are of minor importance; here the pion-pion channels are smaller than proton-nucleon or even pion-nucleon channels. However, the shape of the experimental spectrum is not very well reproduced for 0.3 $\leq$ m $\leq$ 0.5 GeV and the spectrum is also slightly underestimated at invariant masses m $\approx$ 0.3-0.4 GeV. Compared with the results by Li *et al.* [8] - based on an expanding fireball scenario in chemical equilibrium - our $\pi^+\pi^-$ annihilation contribution at low masses is somewhat larger while that from direct $\omega$ decay is slightly smaller.

Since the dilepton yield for 0.3 GeV $\leq$ m $\leq$ 1 GeV is dominated by the $\pi^+\pi^-$ annihilation channel and the electromagnetic form factor of the pion is determined by the properties of the $\rho$-meson [24], in-medium modifications of the latter are expected to affect the dilepton mass spectrum as previously shown in refs. [11, 25]. The actual modifications of the $\rho$-meson in a dense baryonic environment, however, are still a matter of debate. It is thus hoped that low mass dileptons from heavy-ion collisions can shed some light on this issue.

From QCD inspired models [26] or estimates based on QCD sum rules [27, 28] it has been predicted that the $\rho$-meson mass decreases with density. In order to explore the compatibility of such scenarios with the CERES data (as proposed by Li *et al.* [8]), we have performed calculations with a medium-dependent $\rho$-mass according to Hatsuda and Lee [27], i.e.

$$m_\rho^* \approx m_\rho^0 \left(1 - 0.18\rho_B/\rho_0\right) \geq m_u + m_d \approx 14 MeV, \qquad (6)$$

where $\rho_B(t)$ is the actual baryon density during the decay of the $\rho$-meson and $\rho_0 \approx 0.16$ fm$^{-3}$. The dropping of the $\rho$-meson mass is associated with a scalar self-energy of the meson which is determined by the local baryon density; the propagation of the 'quasi-particle' with effective mass $m_\rho^*$ thus couples to the baryon current during the expansion, and the meson becomes 'on-shell' asymptotically due to a feedback of energy from the mean fields, which thus ensures that the total energy of the system is conserved.

The results of this simulation are shown in Fig. 4 in comparison with the CERES data. Again we find the $\pi^+\pi^-$ annihilation to dominate the spectrum for invariant masses m $\geq$ 0.3 GeV, and a significant enhancement of low mass dileptons is obtained. Our computation thus supports the proposal by Li *et al.* [8] that the enhanced dilepton yield might be due to a dropping $\rho$-mass in the medium.

On the other hand, within the vector-dominance model Herrmann *et al.* [29] and Asakawa *et al.* [30] have predicted that the $\rho$-mass does not change much with density, but instead the width of the $\rho$-resonance should increase substantially. Similar, but less pronounced modifications, have been claimed by Chanfray *et al.* [31]. We also follow this assumption as in ref. [11] and explore if such medium effects might actually be seen in the CERES experiment.

In this respect we have performed calculations using a rough fit to the density-dependent pion form factor of Herrmann *et al.* [29] (cf. Fig. 14 of ref. [11]). In order to account for the effect of $\Delta$'s in matter we use an effective density $\rho_N - \rho_\Delta/4$ as pointed out in ref. [32]. The results are displayed in Fig. 5 in comparison to the CERES data [5] and are seen to agree also with the experimental results within the errors bars. Including a collisional broadening of the $\rho$ as calculated by Haglin [33] will further enhance the yield of low mass dileptons. Since the change of the pion form factor in the work of Herrmann *et al.* [29] is entirely due to conventional many-body effects, i.e. the coupling of the $\pi, \rho, \Delta$ and nucleon degrees of freedom, better experimental data and further theoretical studies appear necessary to extract new physics from the CERES data.

In summary, we have studied $e^+e^-$ production in proton and heavy-ion induced reactions at SPS energies on the basis of the covariant transport approach HSD [13] which describes the hadronic processes quite reliably. We have incorporated the contributions from proton-nucleon, pion-nucleon and pion-pion bremsstrahlung, the Dalitz-decay of the $\Delta$, $\eta$ and $\omega$ as well as $\pi^+\pi^-$ annihilation and the direct dilepton decay of the vector mesons $\rho, \omega, \Phi$. It is found that for p + Be and p + Au at 450 GeV the mesonic decays almost completely determine the dilepton yield, whereas in S + Au reactions the $\pi^+\pi^-$ annihilation channel takes over due to the high pion densities achieved. The experimental data taken by the CERES collaboration [5] are slightly underestimated by the calculation when using free form factors for the pion and $\rho$-meson.

Different in-medium effects on the $\rho$-meson have been expected; here we have examined a shift of the $\rho$-mass according to QCD sum rules as suggested by Hatsuda and Lee [27] and the broadening of the $\rho$ according to the approach by Herrmann *et al.* [29]. Although the dropping $\rho$-mass scenario proposed by Li *et al.* [8] is compatible with the CERES data for S+Au, the large error bars in the data can not exclude the more conventional scenario of a broadening $\rho$ spectral function in the medium. To extract direct evidence for chiral symmetry restoration, we need both improved experimental data and further theoretical studies.

We gratefully acknowledge many helpful discussions with A. Drees, B. Friman, K. Haglin, U. Mosel, H. J. Specht and Gy. Wolf. One of us (C.M. Ko) likes to thank U. Mosel for the kind hospitality extended to him during his stay at the University of Giessen under a Humblodt Research Award. His work was also partially supported by the US National Science Foundation under grant No. PHY-9212209 and the Welch foundation under Grant No. A-1110.

# Figure Captions

**Fig. 1:** Comparison of our calculations for the differential dilepton spectra (thick solid lines) with the experimental data [5] (full dots) for p + Be and p + Au at 450 GeV. The individual contributions from the vector meson decays and Dalitz decays as well as bremsstrahlung channels and $\pi^+\pi^-$ annihilation are shown by the thin solid, dashed and dotted lines.

**Fig. 2:** Comparison of our calculations for the proton (dashed line) and $\pi^-$ rapidity distribution (solid line) in comparison to the experimental data [22, 23] for S + Au at 200 GeV/A.

**Fig. 3:** Comparison of our calculations for the differential dilepton spectra (upper thick solid line) with the experimental data [5] (full dots) for S + Au at 200 GeV/A when employing no medium modification of the mesons. The triangles represent the experimental reconstruction from meson decay channels [5] which is compared to the corresponding sum from our calculation (thick solid line denoted by 'cocktail'). The individual contributions from the vector meson decays and Dalitz decays as well as bremsstrahlung channels and $\pi^+\pi^-$ annihilation are shown by the thin solid, dashed and dotted lines.

**Fig. 4:** Comparison of our calculations for the differential dilepton spectra (thick solid line) with the experimental data [5] (full dots) for S + Au at 200 GeV/A when including a shift of the $\rho$-meson mass (6) according to the prediction by Hatsuda and Lee [27]. The individual contributions from the vector meson decays and Dalitz decays as well as bremsstrahlung channels and $\pi^+\pi^-$ annihilation are shown by the thin solid, dashed and dotted lines.

**Fig. 5:** Comparison of our calculations for the differential dilepton spectra (thick solid line) with the experimental data [5] (full dots) for S + Au at 200 GeV/A when including a broadening of the $\rho$-meson according to the calculations by Herrmann *et al.* [29]. The individual contributions from the vector meson decays and Dalitz decays as well as bremsstrahlung channels and $\pi^+\pi^-$ annihilation are shown by the thin solid, dashed and dotted lines.

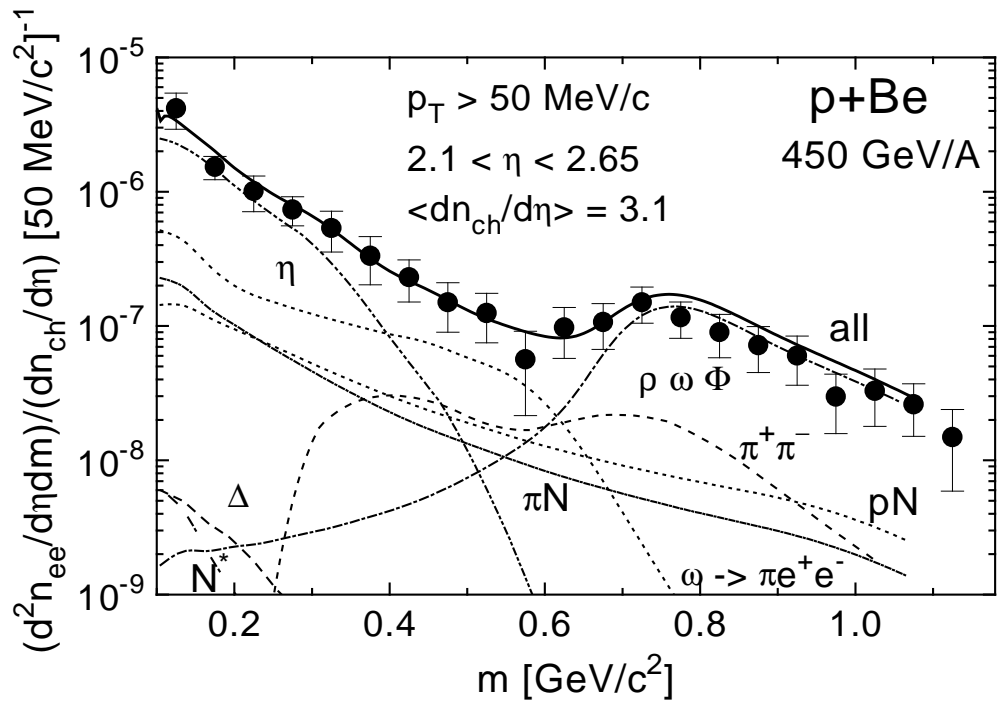

Figure 1a

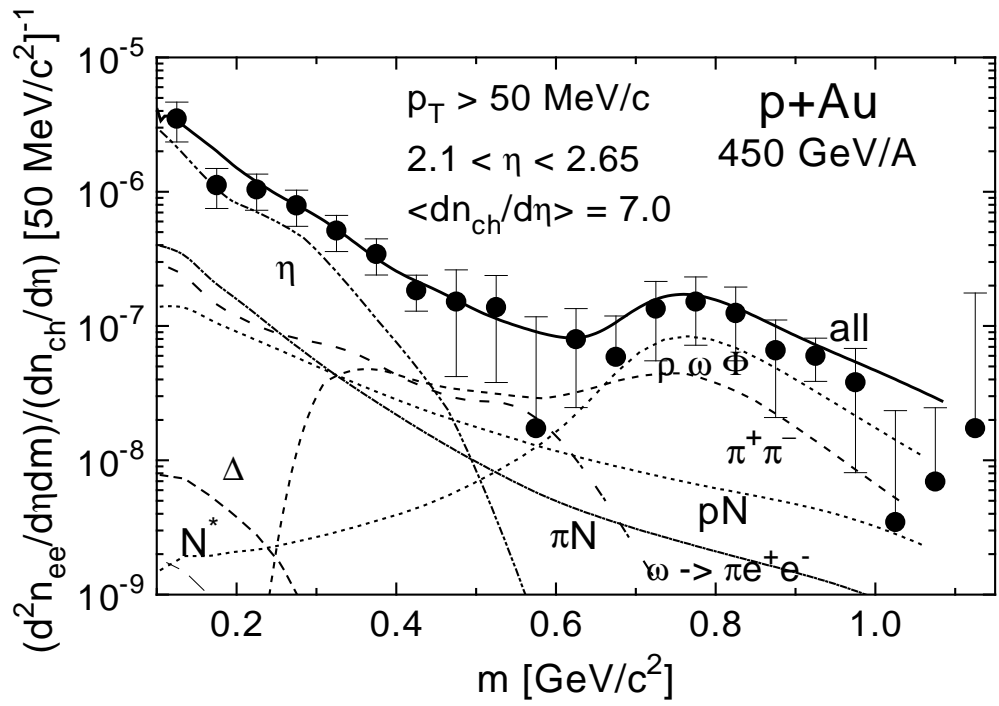

Figure 1b

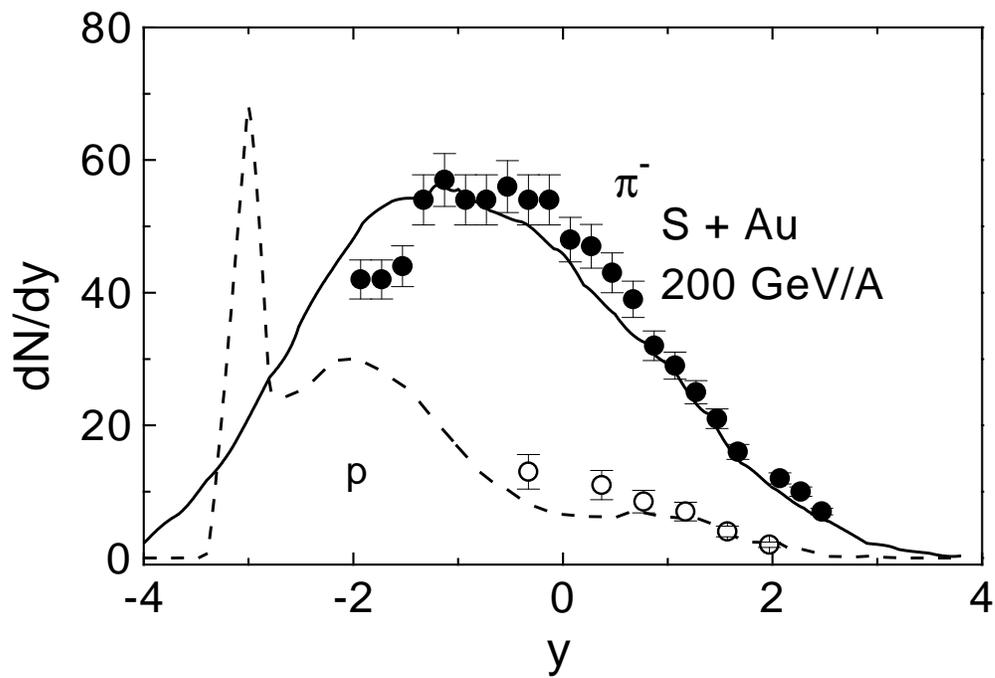

Figure 2

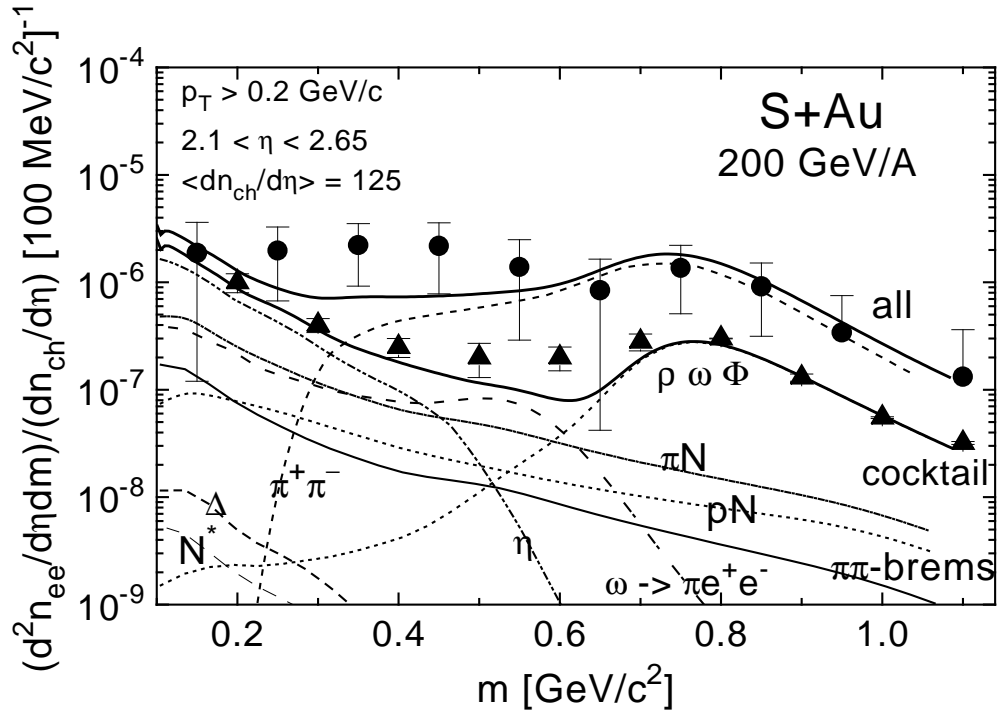

Figure 3

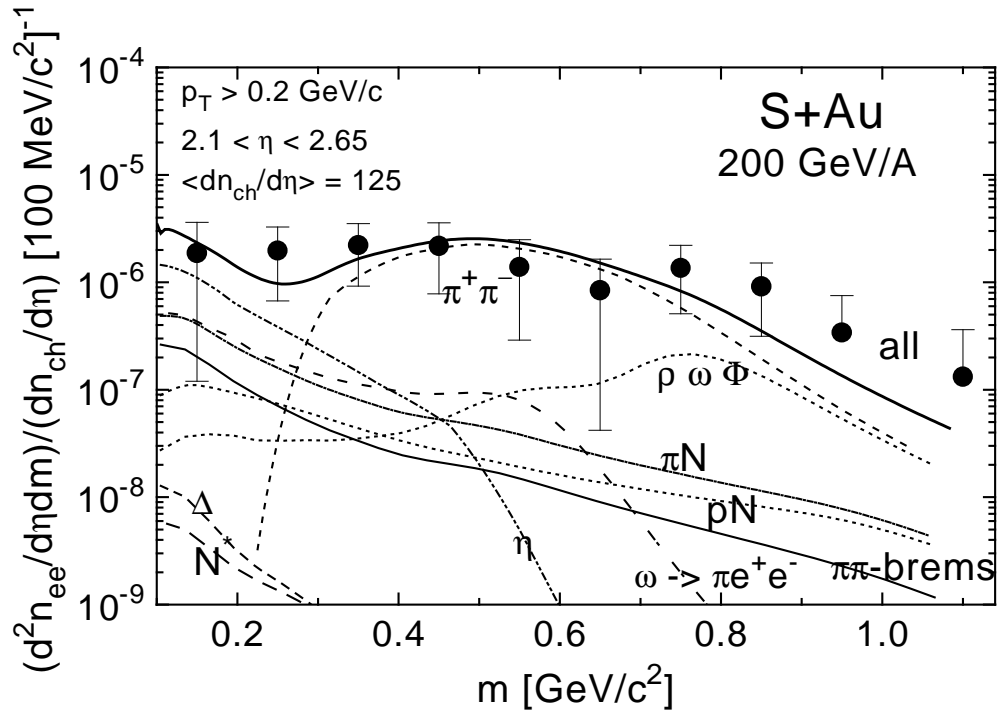

Figure 4

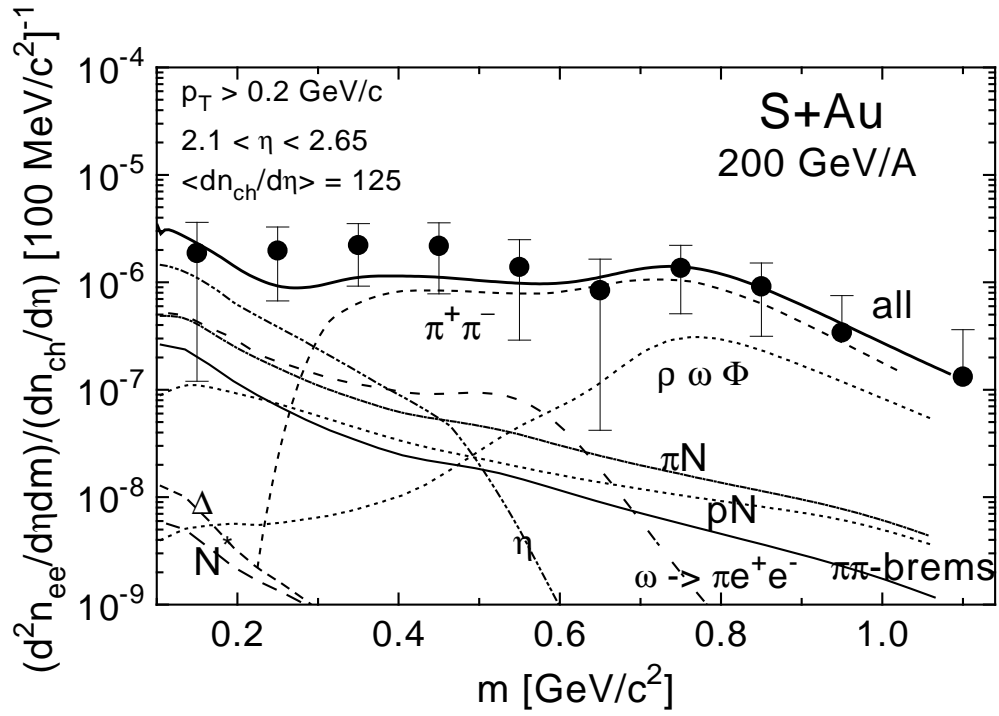

Figure 5